\documentclass[]{article}

\usepackage{jinstpub}
\usepackage{subfigure,dcolumn}
\usepackage{mhchem}
\usepackage{upgreek}
\usepackage{commath}
\usepackage{mathtools}
\usepackage{cases}
\usepackage{lineno}
\usepackage{xcolor}
\usepackage{float}
\usepackage{boxedminipage}
\usepackage{graphicx}
\usepackage{booktabs}
\usepackage{amssymb,bm,mathrsfs,bbm,amscd}

\newcommand{\vect}[1]{\mathbf{#1}}


\author[a]{Qingmin Zhang,}

\author[a,1]{Bangjie Deng,\note{Corresponding author.}}
\emailAdd{dengbangjie@foxmail.com}

\author[a]{Yuanmiaoliang Chen,}

\author[a]{Bochao Liu,}

\author[a]{Shaofei Chen,}

\author[a]{Jinquan Fan,}

\author[b]{Lie Feng,}

\author[b,2]{Haixiao Deng,\note{Corresponding author.}}
\emailAdd{denghaixiao@sinap.ac.cn}

\author[b]{Bo Liu,}

\author[b]{Dong Wang}

\affiliation[a]{Xi'an Jiaotong University, Xi'an, 710049, China}

\affiliation[b]{Shanghai Institute of Applied Physics, Chinese
	Academy of Sciences, Shanghai, 201800, China}

\arxivnumber{1702.06652}

\abstract{
	The free electron laser (FEL), as a next-generation light
	source, is an attractive tool in scientific frontier research
	because of its advantages of full coherence, ultra-short pulse duration,
	and controllable polarization. Owing to the demand of real-time
	bunch diagnosis during FEL experiments, precise
	nondestructive measurements of the polarization and X-ray energy 
	spectrum using one instrument are preferred. In this paper, such an instrument
	based on the electron time-of-flight technique is
	proposed. By considering the complexity and nonlinearity, a
	numerical model in the framework of Geant4 has been developed for
	optimization. Taking the Shanghai Soft X-ray FEL user facility as
	an example, its measurement performances' dependence on the
	critical parameters was studied systematically, and, finally, an
	optimal design was obtained, achieving resolutions of
	$0.5\%$ for the polarization degree and $0.3\ eV$ for the X-ray energy
	spectrum.}

\begin{document}
	

\date{\today}

	\bibliographystyle{JHEP}
	
	\title{Design of a Nondestructive Two-in-One Instrument for Measuring
		the Polarization and Energy Spectrum at an X-ray FEL Facility}

	\maketitle
	\flushbottom

	\section{Introduction}\label{sec:introduction}
	Recently, owing to their irreplaceable advantages of high
	brightness\cite{McNeil2010}, fully transverse
	coherence\cite{Singer2008,Sacla2014}, ultra-short pulse
	duration\cite{Behrens2014} and well-defined
	polarization\cite{Deng2014PRS,Allaria2014Control,Lutman2016},
	several soft and hard X-ray free electron laser (FEL) user facilities, as the next-generation light source, have been designed, under
	construction, or in operation; these include LCLS (SLAC,
	USA)\cite{EmmaP.2010}, SACLA (SPRING8, Japan)\cite{Ishikawa2012}, the
	European XFEL (Hamburg, Germany)\cite{A.S.Schwarz2004a}, FERMI
	(Trieste, Italy)\cite{Walker1996}, and SXFEL (Shanghai,
	China)\cite{Zhaoa}. Currently, SXFEL with a designed wavelength of
	$8.8\ nm$ is under commissioning as an FEL test facility.
	Furthermore, a project upgrading SXFEL to a user facility has
	been launched, in which the electron beam's energy will be
	boosted from $ 0.84 $ to $1.6\ GeV$, to cover the
	water window or even the magnetic window\cite{Song2016}.
	According to the baseline design of the SXFEL user facility, each
	X-ray pulse contains more than $ 10^{12} $ photons with $\sim$ $
	100\ fs $ pulse duration and controllable
	polarization\cite{Zhaoa}.
	
	It is well known that high-resolution measurement of the polarization
	properties and X-ray energy spectrum are strongly demanded in soft X-ray magnetic circular dichroism spectroscopy.
	Meanwhile, the SXFEL user facility requires noninvasive
	measurements of polarization and $ E_{ph} $ with $ 0.5\% $ and $
	0.5\ eV $ precision, respectively.
	However, conventional optical methods, in which transmission
	polarizers or multilayer reflectors are used\cite{Feng2015a}, are no more
	suitable owing to soft X-ray's strong absorption in the interaction
	materials.
	At FERMI, with an X-ray energy range from $22.9 $ to $47.6\ eV$,
	FEL polarization was measured by using three methods: EUV
	light fluorescence, VUV optics, and photoelectron angular distributions \cite{Finetti2014b}. The electron time-of-flight
	(e-TOF) technique is utilized in the third method, so it is routinely
	called \textquotedblleft e-TOF based.\textquotedblright\ It is
	a noninvasive method because only a small fraction of photons
	interact with the rarefied gas in the e-TOF instrument and the others
	go through without any change. 
	However, high resolution can be achieved because of the beam's high intensity. In contrast, the other two methods have more
	serious impact on the photon beam owing to their transmission
	optics and phase retarders\cite{Finetti2014b}. Besides, the e-TOF-based instruments have been successfully applied in a
	Polarization monitor at an X-ray
	FEL\cite{Allaria2014Control,Lutman2016}.
	It is worth mentioning that e-TOF-based X-ray spectroscopy has
	been applied widely for many years and that its time
	resolution has been improved from picoseconds\cite{Nelson2004,Saes2004,March2011} 
	to femtoseconds\cite{Schoenlein2237,Zhang2011,Bressler489}, which has paved the road to precisely measure the photon energy spectrum for
	ultra-short X-ray FEL pulses.
	
	Accordingly, the e-TOF-based method can be utilized to
	simultaneously measure the polarization and energy spectrum of a soft
	X-ray FEL. In this paper, by considering the complexity and
	nonlinearity, a numerical model in the framework of
	Geant4\cite{Agostinelli2003250} is established for simulating
	such a two-in-one instrument, which is based on the e-TOF technique for
	simultaneously measuring the polarization and energy spectrum.
	With the SXFEL user facility taken as an example, its systematic design is devised
	and its optimization is performed. Finally, our study shows that a
	resolution of $0.5\%$ in polarization degree and $0.3\ eV$ in
	X-ray energy spectrum can be obtained with the optimal design.

	\section{Method description}
	\label{sec:Method principles}
	For each X-ray FEL pulse, photoelectrons are produced by X-ray photon photoionizations with target atoms. By measuring their angular distribution
	and drift time, one can simultaneously measure the
	polarization properties and $ E_{ph} $ spectrum of the X-ray FEL. In
	this section, measurement principles will be described in
	detail.
	
	\subsection{Polarization measurement}
	
	The electric vector ($ \vec{E} $) of completely polarized light
	can be expressed in the form 
	\begin{equation}\label{eq:splitfunction}
	\begin{split}
	\vec{E}&=A_x\cos{\omega t}\cdot\vec{e}_x+A_y\cdot\cos(\omega t
	+\Delta\varphi)\cdot\vec{e}_y\\
	&=\underbrace{(A_x+A_y\sin{\Delta \varphi})\cos{(\omega
			t)}\cdot\vec{e}_x+A_y\cos{\Delta\varphi}\cos{(\omega
			t)}\cdot\vec{e}_y}_{\text{\normalsize linearly polarized part}}\\
	&+\underbrace{(-A_y\sin{\Delta \varphi})\cdot\left[\ \cos{(\omega
			t)}\cdot\vec{e}_x+\sin{(\omega t)}\cdot \vec{e}_y\
		\right]}_{\text{\normalsize circularly polarized part}}.
	\end{split}
	\end{equation}
	As shown in Eq. \ref{eq:splitfunction}, completely polarized
	light can be decomposed into linearly polarized light ($ E_{lin} $)
	and circularly polarized light ($ E_{cir} $), which means that linear
	polarized light and circularly polarized light can be used to
	describe the polarization properties of completely polarized
	light. The photoelectron angular distribution of linearly
	polarized photons is different from that of
	circularly polarized or nonpolarized photons; hence, the polarization properties of an X-ray FEL
	can be inferred by using these angular distribution.

	For $\emph{s}$ shells, the photoelectron angular distribution 
	in the plane perpendicular to photon momentum direction
	(in the detection plane) is described precisely enough by the dipole
	approximation\cite{Cooper1993,Manson1982a,Paulus2006}. For a completely polarized X-ray FEL, a normalized
	probability distribution of photoelectrons from $\emph{s}$ 
	shells in the plane perpendicular to the photon direction
	is\cite{Finetti2014b,Allaria2014Control}
	\begin{equation}\label{eq:DifProbDistriburion}
	p(\theta)=\frac{1}{2\pi}+P_{lin}\cdot
	\frac{3\beta}{2\pi(4+\beta)}\cos\left[	2(\theta-\varPsi)	\right],
	\end{equation}
	where $ P_{lin} $ is the linear polarization degree, $ \varPsi $ is the
	polarization angle of linearly polarized photons, $ \theta $ is the angle between the photoelectrons
	momentum and the polarization direction, and $ \beta $ is the dipole
	parameter related to the gas target species.

	However, for $\emph{p}$ shells, the differential photoelectric
	cross section cannot be described by the electric dipole
	approximation, which results in complexity for data analysis.
	However, electrons from $\emph{p}$ shells can be excluded by
	the longer time of flight, which is related to their lower kinetic energy.

	According to Eq. \ref{eq:DifProbDistriburion}, $ p(\theta) $
	can be obtained by fitting an electron angular distribution
	with $ p(\theta)=A+B\cdot\cos{[2(\theta-C)]} $. Then, the linear
	(circular) polarization degree $ P_{lin} $ ($P_{cir}$) would be
	\begin{linenomath}
		\begin{numcases}{}
		P_{lin}=\dfrac{B}{A}\cdot\dfrac{3\beta}{4+\beta},\label{eq:Plin}&\\
		P_{cir}=\sqrt{1-P_{lin}^2}\label{eq:Pcir}&
		\end{numcases}
	\end{linenomath}
	and polarization angle would be $ \varPsi=C $.
	\subsection{$ E_{ph} $ spectrum measurement}
	
	Photon energy can also be derived by the
	time of flight, since the relation between photoelectron
	energy $( E_{e} )$ and photon energy $ (E_{ph}) $ is
	\begin{equation}\label{eq:binding energy}
	E_{ph}=E_e+E_b,
	\end{equation}
	where $ E_b $ represents the electron's binding energy for a given
	shell, which varies from gas to gas\cite{TRZHASKOVSKAYA2001,trzhaskovskaya2002photoelectron}.
	By measuring the electron's flight time for a given drift length
	$L $, $ E_{ph} $ can be determined with
	\begin{equation}
	\label{eq:Photon Energy}
	E_{ph}=\dfrac{m_eL^2}{2{\Delta t}^2}+E_b,
	\end{equation}
	where $ \Delta t $ is the photoelectron flight time and $
	m_e $ is the electron's rest mass.
	
	As mentioned above, $ E_b $ varies with target gases. However,
	because the angular distribution of Auger electrons is different from
	that of photoelectrons ($ E_{k,ph} $), gas selection should
	avoid overlap with the kinetic energy of Auger electrons ($
	E_{k,Auger} $), which is
	\begin{equation}
	E_{k,Auger}=E_i-E_j-E_k,
	\end{equation}
	where $E_i$, $E_j$, and $ E_k $ represent the electron binding
	energies for shell $ i $, $ j $ and $ k $, respectively. These
	properties can be found in References \cite{Paulus2006} and
	\cite{Yamazaki2007}.
	
	By making use of pulse waveform analysis and earliest arrival time
	measurement, one can obtain the flight time spectrum for photoelectrons, from which $ E_{ph} $ spectrum can be derived
	according to Eq. \ref{eq:Photon Energy}.
	
	\section{Overall design}
	\label{sec:Overall design}
	\begin{figure*}[htp]
		\centering
		\includegraphics[width=\textwidth]{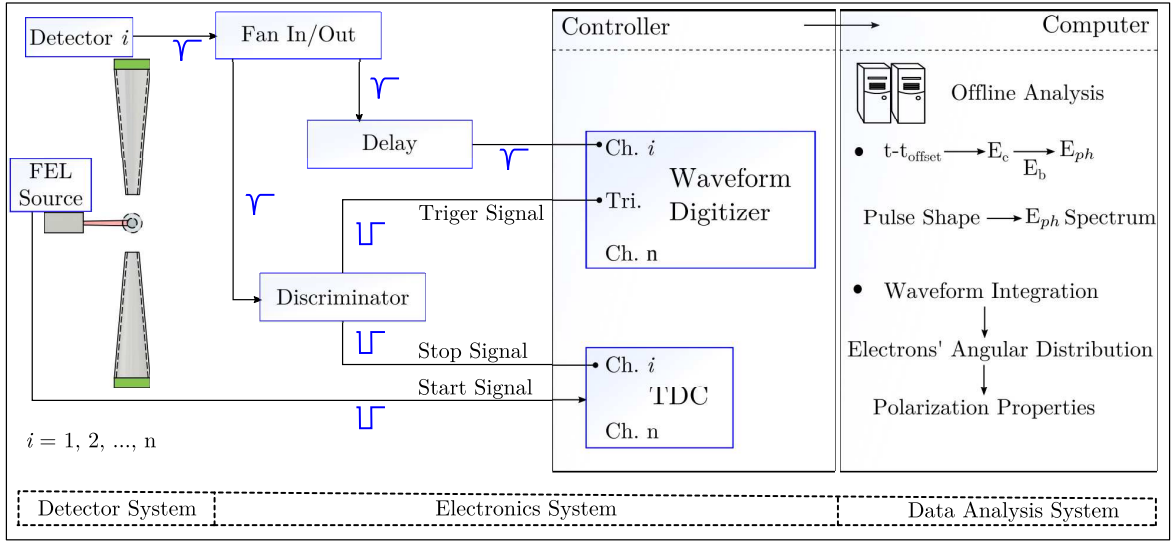}
		\caption{Block diagram of the apparatus.}
		\label{fig:Data analysis arrangement}
	\end{figure*}
	According to the physical principles described above, 
	the photoelectron angular distribution can be measured by 
	placing detectors in the detection plane and their energy
	spectrum can be obtained by using the time of flight. The instrument
	is designed to contain three subsystems: a detector, electronics,
	and data analysis, as demonstrated in Fig. \ref{fig:Data
		analysis arrangement}.

	The detector system is composed of a differential vacuum system and
	a given number of identical e-TOF detectors for photoelectron
	generation, drifting, and detection. The gas target is located at the
	center of the instrument for photoelectric reaction, being
	uniformly surrounded by e-TOF detectors. Each e-TOF detector is
	an electron multiplier equipped with a collimation tube.
	Micro-channel plates (MCPs) are preferred as electron multipliers owing 
	to their timing resolution of hundreds of picoseconds because of their thinner micro-channels and better
	response linearity because of their numerous independent micro-channels and electron dispersive arrival at the MCP in space.
	In addition, circular truncated cone collimation tubes were designed
	for better placement. Following the X-ray FEL beam's arrival,
	photoelectrons are generated in the central volume, drift through the
	collimation tubes, and are finally detected by electron detectors.
	Given statistical errors, the pressure in the central volume should be fairly high for generating enough photoelectrons, while
	its surrounding volume should be at very low pressure to minimize
	electron loss and scattering during drifting and to
	satisfy the detectors' working pressure requirement. Accordingly,
	a differential vacuum system (DVS) was adopted. A small
	difference between the target gas and that of the drifting environment is
	preferable to avoid complicated DVS design.

	The electronics system was designed to record the flight time and
	signal waveform of photoelectrons by using a time-to-digital
	converter (TDC) unit and a waveform digitizer for all detectors, respectively.
	For each X-ray FEL pulse, the TDC is triggered by a start signal supplied by
	The timing system of the FEL source and stopped by a signal from a
	discriminator. Meanwhile, the stop signal is also
	transferred to the waveform digitizer for signal shape
	recording. A proper delay for the detection signal transferred
	to the waveform digitizer is introduced for synchronization in
	order to record the full signal shape. The signal
	shape can be integrated to obtain the signal charge, which is
	proportional to the number of photoelectrons. Then, the
	photoelectron angular distribution is obtained to conclude attainment of the
	polarization properties. By combining the signal shape and flight
	time, the photoelectron flight time spectrum can be determined.

	The data analysis system was designed to analyze waveform time
	series data, from which the photoelectron flight time spectrum
	can be derived by using pulse shape analysis methods.
	Furthermore, the $ E_{ph} $ spectrum can be obtained according to
	Eq. \ref{eq:Photon Energy} after an offset time correction. It is
	reasonably assumed that there is no saturation effect for the
	detector's response owing to the moderate number of detected
	electrons and that the pulse shape for each electron is essentially the same. 
	Therefore, the signal's integration charge for each channel ($ Q_i,\
	i=1,2, \dots,16$) is proportional to the number of detected
	photoelectrons. Then the X-ray FEL's polarization properties
	($P_{lin}$ and $P_{cir}$) can be obtained by fitting $ Q_i $ with $
	p(\theta)=A+B\cos{2\left[ (\theta+C) \right]} $ and applying Eq.
	\ref{eq:Plin} to Eq. \ref{eq:Pcir}. Based on the assumptions
	mentioned above, a linear model was proposed to analyze the signal
	waveform to obtain the flight time spectrum. The sampled
	signal pulse $ P(i\Delta t_s)\ (i=1,2,\dots,n) $ with sampling
	time interval $ \Delta t_s $ can be described as
	\begin{equation}\label{sampled shape}
	P=\left[P_1, P_2, \dots, P_i,\dots, P_n-1,P_n\right]
	\end{equation}
	Because $ \Delta t_s $ for the waveform digitizer is much larger than
	the realistic interval of electron arrival time, the sampled signal
	should be interpolated to acquire a smaller time interval $ \Delta
	t_i $ for decomposition. The interpolation signal pulse shape $
	I(j \Delta t_i),(j=1,2,\dots,m) $ can be obtained by 
	interpolating the sampled signal $ P(i\Delta t_s) $ in a small
	interpolation time interval $ \Delta t_i $.	Obviously, $ m\geq n
	$ is required and $ n\Delta t_s $ is much longer than the typical
	width of the whole signal. Commonly, the amplitude of a single
	photoelectron's charge signal ($ S_{0,j} $) with time interval
	$ \Delta t_i $ in a simulation time $t= j\Delta t_i\ (j=1, 2,
	\dots, m) $ can be described by the log-normal distribution $ f(t)
	$ in the form 
	\begin{equation}\label{eq:SignalPulse}
	f(t)=e^{-\dfrac{1}{2}\cdot\left(\dfrac{\log(t/\tau)}{\sigma_s}\right)^2},
	\end{equation}
	where $ \tau$ and $\sigma_s $ define the signal's location and scale,
	respectively\cite{Jetter2012}.	Therefore, the signal's pulse shape $
	(\vect{S_0}) $ for a single electron without delay can be
	described digitally as follows:
	\begin{linenomath}
		\begin{numcases}{}
		\vect{S_0}=\left[S_{0,1},\ S_{0,2},\ \dots,S_{0,j},\ \dots,\
		S_{0,m} \right]^T,\label{eq:S0}&\\		S_{0,j}=f(j\Delta t_i)\qquad
		(j=1, 2, \dots, m).\label{eq:S0m}&
		\end{numcases}
	\end{linenomath}
	Thus, the pulse shape for a single electron with a delay of $
	k\Delta t_i\ (k=0, 1, 2, \dots, m-1) $ can be expressed
	digitally as 
	\begin{linenomath}
		\begin{numcases}{}
		\vect{S_k}=\left[S_{k,1},S_{k,2},\dots,S_{k,j},\dots,S_{k,m}
		\right]^T,&\\
		S_{k,j}=f((j-k)\Delta t_i)\qquad (j=1, 2, \dots,
		m).\label{eq:Stm}&
		\end{numcases}
	\end{linenomath}
	Then, signal matrix for a series of single electrons with a delay
	$ k\Delta t $ can be written as
	\begin{equation}
	\vect{S}_{m\times m}=\left[\vect{S_0}, \vect{S_1}, \dots,
	\vect{S_{m-2}}, \vect{S_{m-1}} \right]
	\end{equation}
	In addition, the number of electrons for each bin can be expressed as
	\begin{equation}
	\vect{C}_{m\times 1}={[C_0,C_1,\dots,C_{m-2},C_{m-1}]^T}.
	\end{equation}
	Thus,
	\begin{equation}
	\label{eq:output signal}
	\vect{I}=\vect{S}\cdot \vect{C}+\mathbf{e},
	\end{equation}
	where $\mathbf{e}$ is the error term resulting from noise. Therefore, the
	number of electrons for each timing bin is obtained by solving
	Eq. \ref{eq:output signal}, and, namely, the electron's flight time
	spectrum can be obtained. A least-squares method is chosen for
	optimization in solving for $\vect{C}$ in Eq. \ref{eq:output signal}.
	The estimated number of photoelectrons, $\vect{\hat{C}}$, can be
	obtained by using various optimization algorithms with constraints
	($C_j\geq0,\ j=0,1,\dots,m-1$), such as the conjugate gradient
	method\cite{Hestenes1952}, the Broyde--Fletcher--Goldfarb--Shanno 
	method\cite{Introduction1989a}, etc. Finally, the $ E_{ph} $ spectrum
	can be derived by using Eq. \ref{eq:Photon Energy}.
	
	\section{Numerical modeling}
	\label{sec:Numerical model}
	
	Because the instrument's performance is influenced by many
	factors and photoelectron interactions with the target gas cannot be well
	described analytically, it is impossible to optimize the design
	analytically. Hence, a numerical simulation model will be
	established in the framework of Geant4, which is a Monte Carlo
	nuclear physics simulation software package\cite{Agostinelli2003250}. To simulate polarized photon interactions with the target gas,
	the Livermore Polarized Physics Model with G4EMLOW-6.48
	data\cite{Champion2009} was used and a proper energy threshold of
	100 eV was set for tracking all particles. The major processes included in the
	simulation were the X-ray photoelectric process, electron
	ionization, and scattering.
	
	\subsection{Model description}
	To simulate the detector physics, a detector model
	was established that included the geometrical construction, the
	corresponding physics, and incident photons properties.

\begin{figure*}[htp]
		\centering
		\includegraphics[width=140mm]{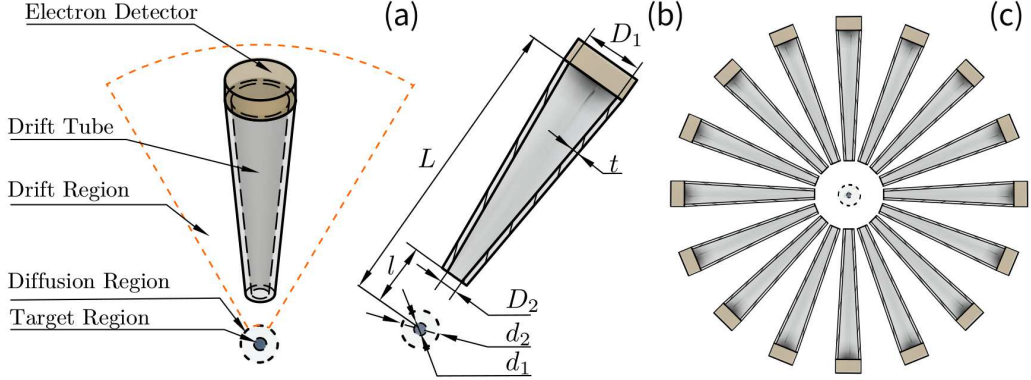}
		\caption{Sketch of e-TOF detectors and gas environment. (a)
			An e-TOF detector and gas regions of the spectrometer. (b)
			Critical parameters in design. (c) Overall layout view of e-TOF
			detectors.}
		\label{fig:Sketch of Spectrometer}
	\end{figure*}

	In the numerical model, instrument's geometry was simplified
	appropriately, as demonstrated in Fig. \ref{fig:Sketch of
		Spectrometer}{(a)}. Electron multipliers and collimating tubes
	were set in the shape of thin cylinders and hollow cone shells,
	respectively. The critical parameters are labeled in Fig.
	\ref{fig:Sketch of Spectrometer}(b) and an overall layout view of the instrument is demonstrated in Fig.
	\ref{fig:Sketch of Spectrometer}(c).
	
	The geometry of the differential vacuum system was omitted and the
	gaseous environment was divided into three parts according to the
	pressure magnitude: the target region, the diffusion region, and the drift region. The target region, a high-pressure cylindrical region, was
	located at the center of the instrument, where gas was injected and
	photoelectrons were produced. Since the FEL pulse duration is
	ultra short, the pressure in the target region was considered to be
	stable during a single pulse. Generally, a target gas with a
	higher $ E_b $ (such as $ O_2 $) means a better energy resolution
	of $ E_{ph} $ for a given drift length. However, to test
	the model in a moderate condition, $ { N_2}\ ( E_b = 403\ eV )$ was
	adopted. The drift region, a steady low-pressure region for
	electron drifting, is used to reduce electron scattering and to
	satisfy the electron multipliers' working environment (in which the pressure
	should be $<$ $0.1\ Pa$). To simulate the pressure
	transition region between the target region and the drift region, a
	diffusion region with 16 radius bins in linear gradient pressure
	was adopted in the numerical simulation. Additionally, we also considered
	terrestrial magnetism, which was set to be
	perpendicular to the detection plane, as it bends electrons
	maximally for such a setup.
	
	The incident polarized X-ray photons were sampled according to
	the start-to-end FEL simulation under typical SXFEL working
	conditions, in which the FEL's average energy is $\sim$ $ 621\ eV $
	with a narrow bandwidth of $\sim$ $ 0.8\ eV $\cite{Song2016}. The
	photon direction was set to be perpendicular to the detection
	plane. In terms of photon polarity, completely polarized
	photons were simulated by the combination of linearly polarized light
	and circularly polarized light photons with a given $ P_{lin} $ (determined
	by the FEL working conditions). The linearly polarized photon polarization
	angle was set to be $ \varPsi $, and the $ \vec{E}_{cir} $
	direction for each circularly polarized photon was sampled from
	a uniform distribution in $ [0,2\pi] $. Meanwhile, the photon beam's
	transverse distribution was considered as a normal distribution.
	Usually, the beam diameter ($ D_b $) was defined as the place where
	photon intensity is reduced to $ 1/e^2 $ of its
	maximum\cite{Siegman:98}, which was calculated as $ 4 \sigma $
	of the normal distribution.
	
	\subsection{Simulation}
	The electronics signals were
	generated according to the time when photoelectrons were detected and a typical pulse shape of an MCP detector. The signals were digitized by using a common sampling time of a waveform digitizer. Details are described below and
	simulation results are presented.

	Because there are $>$ $ 10^{12} $ photons in a single laser
	pulse, a variance reduction technique of cross-section bias for
	only the photoelectric process was adopted. The cross section for the
	photoelectric process was increased by a factor of $ 10^5 $, while the other cross sections remained unchanged. Therefore, the number of total
	sampling events was reduced significantly, while the number of effective events
	remained the same.
	
	For each X-ray FEL pulse, once photoelectrons were detected, their charge
	and drift time were recorded. The detectors' simulated signals
	were produced according to Eq. \ref{eq:SignalPulse} with proper
	shape parameters ($ \tau$  and$\sigma_s $). The discriminator, fan
	in/out unit, delay cables, and a TDC unit (measuring the earliest
	electron's arrival time) were omitted because the e-TOF signal shape
	can be obtained directly in the simulation. By adopting the method
	mentioned above, the angular distribution can be derived by
	integration of the signal shape. Meanwhile, the $ E_{ph} $ spectrum
	was derived by using pulse shape analysis.
	
	\begin{figure}[htp]
		\centering
		\includegraphics[width=85mm]{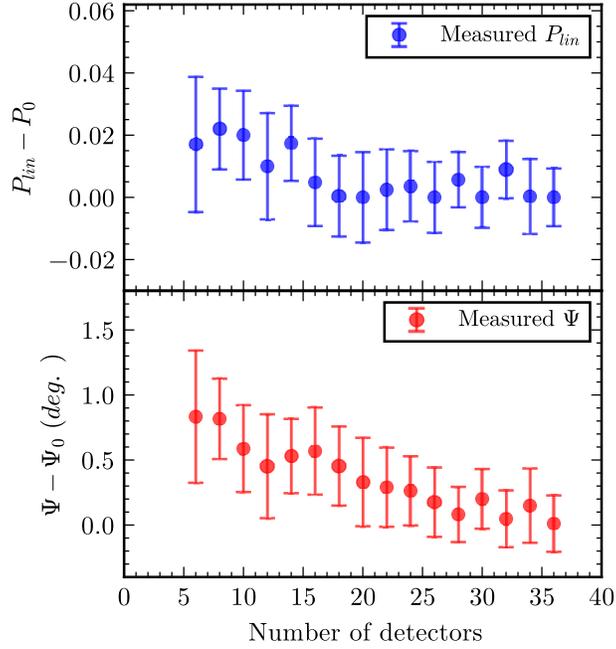}
		\caption{Performance of polarization measurement as a function
			of the number of detectors.}
		\label{fig:NumberofDetectors}
	\end{figure}
	
	The effect of terrestrial magnetism can be corrected by various
	methods, such as magnetic shielding and imposition of a reversed magnetic field.
	Therefore, it was omitted in this simulation and the dependence
	on it will be studied later. Key parameters of the baseline design
	are listed in Table \ref{tab:Critical parameters' value}. It is
	worth mentioning that the detectors are uniformly deployed for
	measuring any polarization angle and that the number of detectors
	is determined by a balance of cost, complexity, and performance
	dependence analysis. Because this instrument's primary goal is to
	measure polarization precisely, the performance of the polarization
	measurement was analyzed by varying the number of detectors with
	the other baseline parameters being exactly same as those of Table
	\ref{tab:Critical parameters' value}. As shown in
	Fig. \ref{fig:NumberofDetectors}, one can see that the performance
	improvement for the polarization measurement is insensitive to the
	number of detectors when the number of detectors is $>$16. Hence, the number of detectors is fixed at 16, and the number of detectors (16) used in the previous setup \cite{Allaria2014Control} is also confirmed as an optimal choice by the simulations conducted in this paper.
	
\begin{figure*}
		\centering
		\includegraphics[width=130mm]{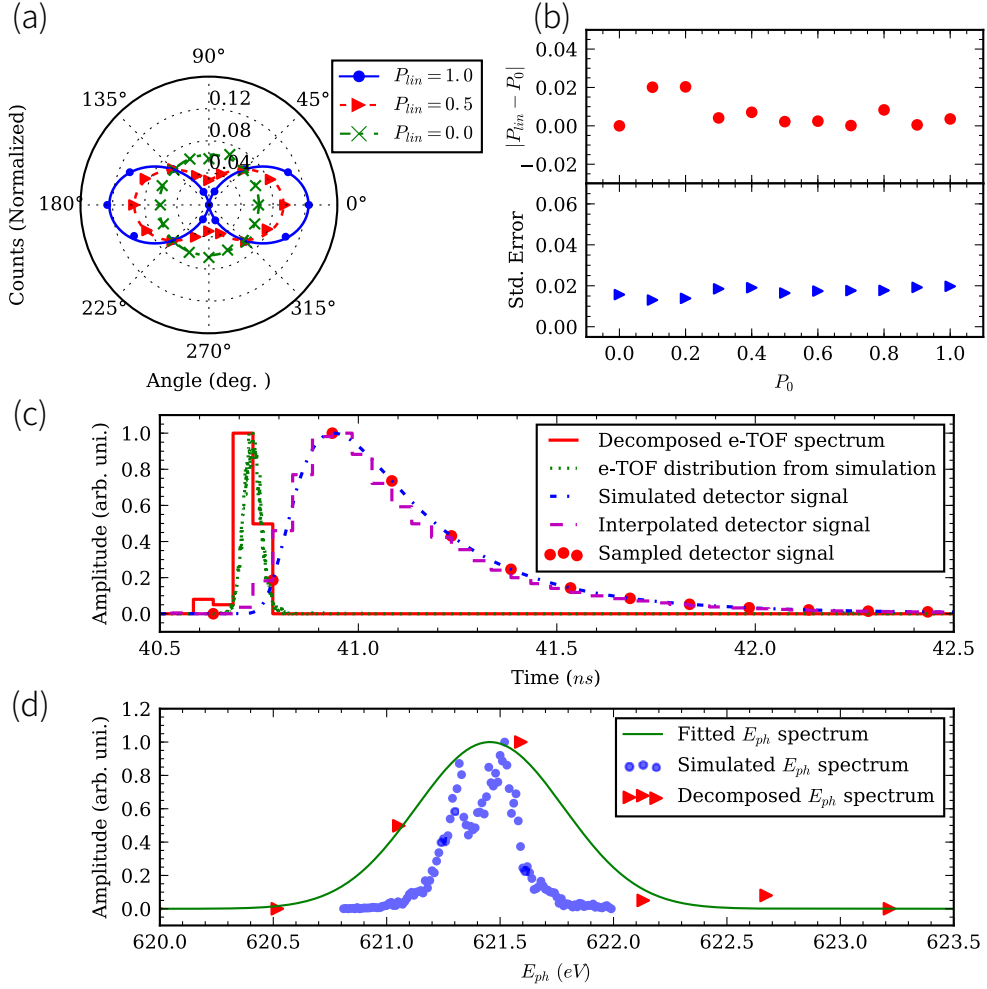}
		\caption{Simulation results for the baseline design. (a) Simulated photoelectron angular distribution for an X-ray FEL with a linear polarization angle of zero degrees in the polar coordinate system as a function of $
			P_{lin} $. (b) Absolute
			deviation of polarization measurement for different linear
			polarization fractions $ P_0 $ in the simulation. (c) Decomposed e-TOF spectrum from detector signals, which is compared with a realistic 
			electron TOF distribution. The simulated e-TOF signal (dashed and
			dotted curve) was sampled by the waveform digitizer in $ \Delta t_s=150\ ps
			$ and interpolated with $ \Delta t_i=50\ ps $. After offline
			analysis, the decomposed e-TOF spectrum (solid curve) is capable of
			describing the e-TOF distribution from the simulation
			(dotted curve). (d) Comparison between the decomposed energy spectrum
			(fitted by a normal distribution) and the $ E_{ph} $ spectrum from the
			start-to-end simulation of an X-ray FEL.}
		\label{fig:baseline}
	\end{figure*}

	In addition, the performance for the baseline design is shown in
	Fig. \ref{fig:baseline}, which verifies the numerical model. 
	
	\begin{table}[htp]
		\centering
		\caption{\label{tab:Critical parameters' value}Critical
			parameters of the baseline design}\begin{flushright}
			
		\end{flushright}
		\begin{tabular}{|l|l|l|}
			\hline
			Parameter  		&Value 				&Description\\
			\hline
			
			$ E_{ph,av} $	&$621.4\ eV$	&Average energy of a typical SXFEL
			pulse\\
			$E_{ph,band}$&$0.9\ eV$&Bandwidth of a typical SXFEL pulse\\
			$ N $				& $ 16 $			&Number of detectors\\
			Gas					& $ N_2 $			&Target gas species\\
			$ E_b $				& $ 403\ eV $		&Binding energy of $ 1s $	orbit\\
			$d_1$				& $5\ mm$			&Diameter of the target region\\
			$d_2$				& $10\ mm$			&Diameter of the diffusion region\\
			$D_b$				& $ 0.1\ mm $		&Average diameter of the laser beam\\
			$ L $				& $ 350\ mm $		&Drift distance\\
			$ l $				& $ 30\ mm $		&Collimating tubes' offset\\
			$D_1$					&$ 27\ mm $			&Diameter of electron multipliers\\
			$D_2$					&$3.24\ mm$			&Diameter of collimation tubes' front
			end\\
			$B$					&$ 0\ gauss $			&Terrestrial magnetic field
			intensity \\
			$\tau$&$200\ ps$ &Location parameter of detector's response
			for an electron \\
			$\sigma_s$&$0.7$ &Scale parameter of detector's response for
			an electron\\
			$t_s$				&$ 150\ ps $			&Sampling time interval\\
			$ t_i $ &$ 50\ ps $	&Interpolation time interval\\
			\hline
	\end{tabular}
	\end{table}

	\section{Dependence analysis}
	\label{sec:Dependence Analysis}
	
	Resolutions of $ 0.5\%$ for the polarization degree measurement and
	$0.5\ eV$ for the $ E_{ph} $ spectrum measurement are required by the
	SXFEL user facility, so the critical parameters' influences on the
	polarization and $ E_{ph} $ spectrum measurements are studied in
	this section.

	\subsection{Polarization measurement}
	The precision of the polarization measurement is related to the number
	and angular distribution of detected photoelectrons. Potential
	critical factors might include the gas pressure for both the target
	region and the drift region, electron drift length, detector size,
	and the residual terrestrial magnetic field. To examine
	their influence, polarization measurement performance will be
	evaluated by using the absolute deviation and standard error of $
	P_{lin} $ for an incident FEL pulse with linear polarization
	degree $ P_0=1 $.

	For the target region, a gas target with a high pressure may mean a larger
	photoelectron yield and a smaller statistical error term.
	However, at the same time, it also has a big impact 
	photoelectron drift, resulting in a nonconvergent
	measurement. Accordingly, the pressure needs to be optimized, and the related
	simulation results are shown in Fig. \ref{fig:Reaction Region
		Pressure}, indicating an optimal $ P_t $ in the range from $ 0.1
$ to $ 1.0\ Pa $.
	
\begin{figure}[htp]
		\begin{minipage}[t]{0.4\textwidth}
			\centering
			\includegraphics[width=70mm]{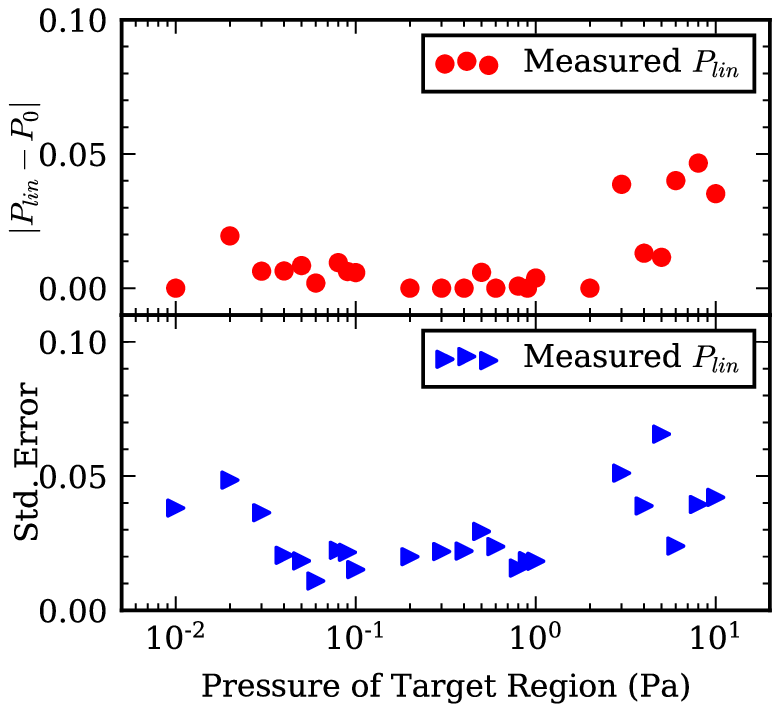}
			\caption{Deviation of polarization measurement with
				changing pressure in the target region.}
			\label{fig:Reaction Region Pressure}
		\end{minipage}
		\hspace{60pt}
		\begin{minipage}[t]{0.4\textwidth}
			\centering
			\includegraphics[width=70mm]{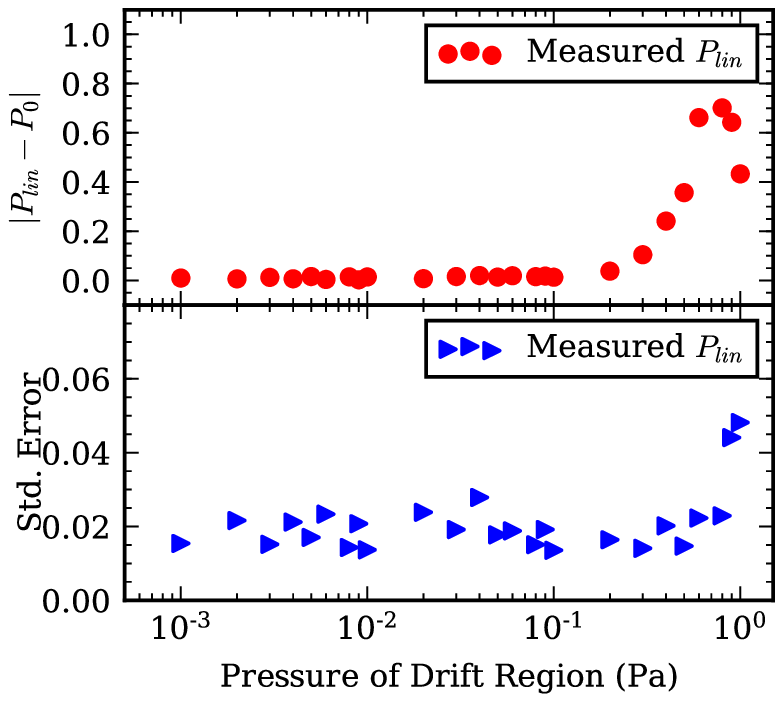}
			\caption{Deviation of polarization measurement with
				changing pressure in the drift region.}
			\label{fig:Background Pressure}
		\end{minipage}
	\end{figure}

	For the drift region, a gas environment with high pressure prevents photoelectron
	drift. As shown in Fig. \ref{fig:Background Pressure}, when $
	P_d < 0.1\ Pa $, deviations of the measured $ P_{lin} $
	change insignificantly and satisfy the corresponding design
	requirements. 
	
\begin{figure}[htp]
		\begin{minipage}[t]{0.4\linewidth}
			\centering
			\includegraphics[width=70mm]{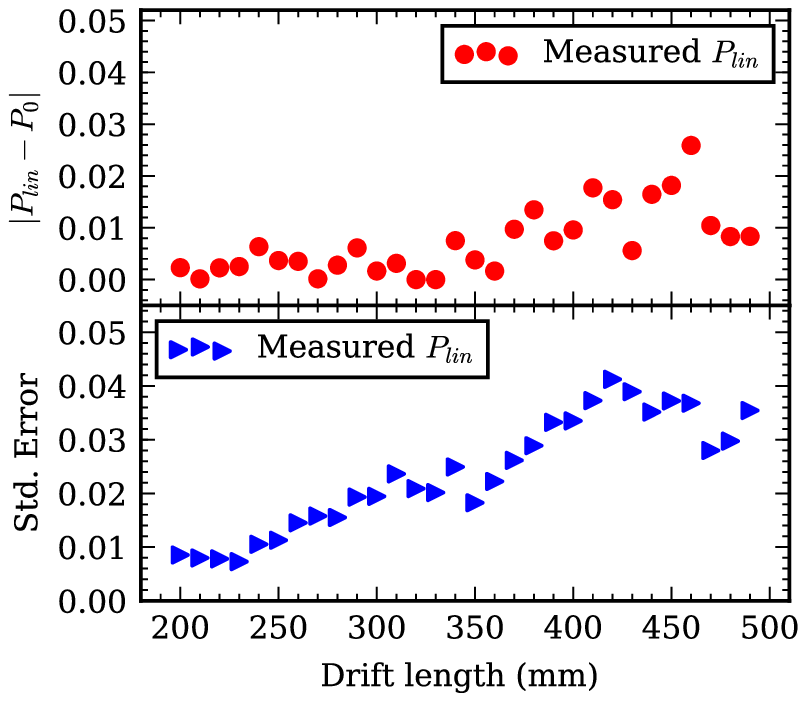}
			\caption{Deviation of polarization measurement with
				changing drift distance.}
			\label{fig:DriftDistance-a}
		\end{minipage}
		\hspace{60pt}
		\begin{minipage}[t]{0.4\textwidth}
			\centering
			\includegraphics[width=70mm]{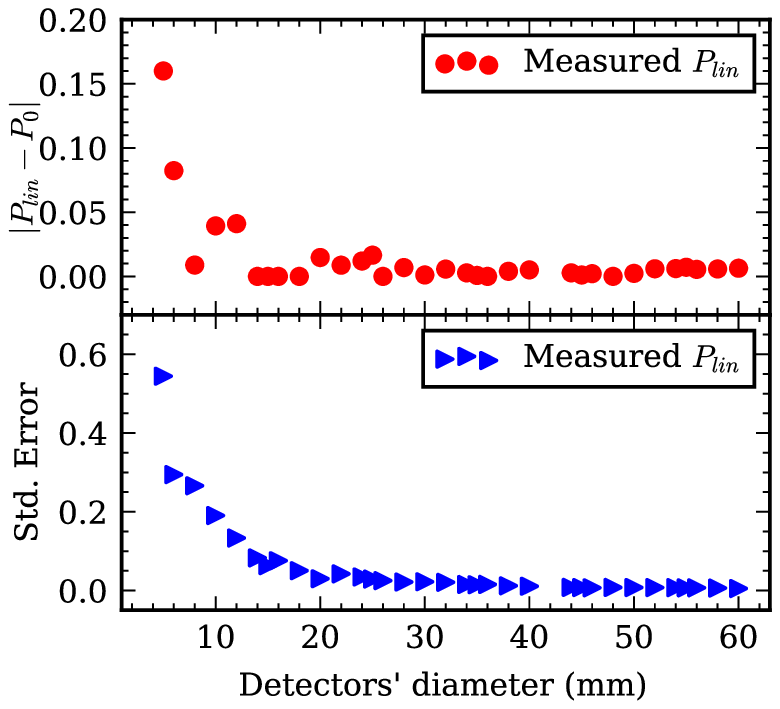}
			\caption{Deviation of polarization measurement with
				changing detector size.}
			\label{fig:Detector Diameter}
		\end{minipage}
	\end{figure}

	A longer drift length indicates a larger statistical error owing to the smaller acceptance angle. As can be inferred from Fig.
	\ref{fig:DriftDistance-a}, as drift distance increases, $|\Delta
	P_{lin}|$ changes slightly. Therefore, an acceptable drift length
	might be from $200 $ to $ 400\ mm $ to satisfy the design
	requirement.
	
	Electron detectors with smaller size might lead to greater
	statistical error owing to their small acceptance angle, while larger
	detectors might cause error term because of the unsuitable assumption of the
	detection plane and inaccurate positioning. The result is
	demonstrated in Fig. \ref{fig:Detector Diameter}. From the
	result, detected photoelectrons can be regarded as a point
	source and to be within the detection plane when the detectors'
	diameter ($D_2  $) is in the range from $ 5 $ to $ 60\ mm $
	and the drift length $ L $ is $ 350\ mm $. Meanwhile, $ D_2 $ should
	be $>$ $30 mm$ (typically 42 mm for commercial
	products) to comply with the design requirement.
	
	\begin{figure}[htp]
		\centering
		\includegraphics[width=140mm]{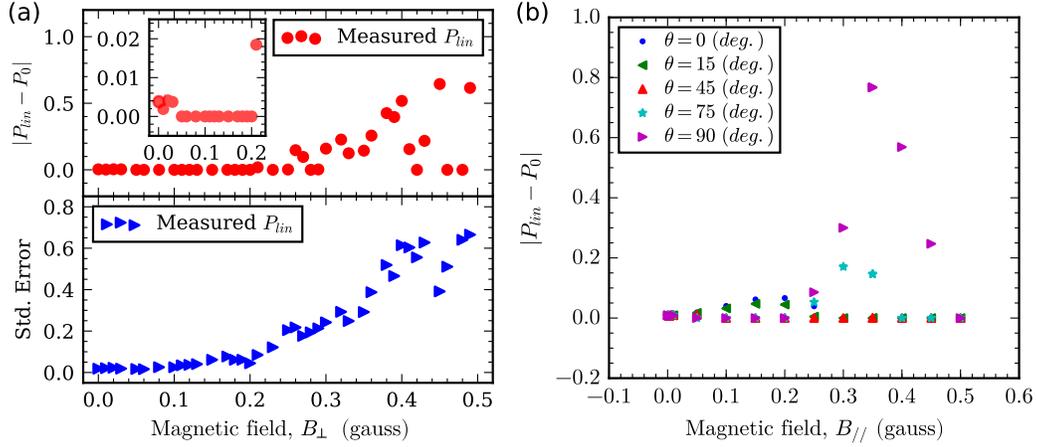}
		\caption{Deviation of polarization measurement with changing residual Earth magnetic field in the direction (a) perpendicular to the detection plane and (b) parallel to the detection plane, respectively.}
		\label{fig:Remnant of Earth Magnetism}
	\end{figure}
	
	The terrestrial magnetic field bends the drift path of the photoelectrons,
	resulting in deviations for measurement. Such deviations can be
	avoided by magnetic shielding or imposing a reversed magnetic field;
	however, how weak the residual magnetic field should be is a
	question to be answered by us. The magnetic field in any direction can be decomposed into orthogonal directions : perpendicular ($B_{\perp}$) and parallel ($B_{//}$) to the detection plane. The influence of the perpendicular component is same for photoelectrons at any direction in detection plane, while the influence of the parallel component is different for photo-electrons in different directions. So the dependence of the polarization measurements on the residual terrestrial magnetic field for these two directions was
	studied. As shown in Fig. \ref{fig:Remnant of Earth Magnetism}, a maximum residual magnetic field of $\sim$ $ 0.2 $ gauss in perpendicular direction for any linear polarization angle and in parallel direction for polarization angle of about $45^\circ$ is small enough to ignore its influence. For $B_{//}$ with other linear polarization angles, a more strict magnetic shielding is demanded to improve measurement precision. The terrestrial magnetic declination and the magnetic inclination at SXFEL site are $ -5^\circ 52' $ and $ 46^\circ52'$, respectively \cite{magnet-declination}. Additionally, the designed beam direction and
	linear polarization of SXFEL are about $ 11^\circ $ south of west and almost horizontal in detection plane, respectively. Based on these facts, further simulations were performed to evaluate the influence of residual magnetic field for real situation, showing a maximum residual magnetic field of $\sim$ $ 0.2 $ gauss can meet the requirement of measurement precision and higher magnetic shielding is strongly required for other linear polarization angles.

	\subsection{$ E_{ph} $ spectrum measurement}
	For the $ E_{ph} $ spectrum
	measurement, longer signal duration and smaller
	sampling and interpolation time intervals are preferred, both of which mean higher energy
	resolution. A longer signal duration can be achieved by a longer
	drift length $(L)$ and by using a target gas with a higher binding energy.
	The sampling time interval $\Delta t_s$ is often determined by the
	maximum  sampling rate of the waveform digitizer. Larger $\Delta t_s$
	means worse precision of sampled signals, which leads to distortion in later interpolation. Moreover,
	the interpolation time interval $\Delta t_i$ should be chosen
	according to the full width at half maximum, because two overlapped signal peaks can
	be mistaken as one peak. In the following, we study the following key parameters:
	species of target gas, drift length $L$, and sampling and
	interpolation time intervals ($\Delta t_s$\ and $\Delta t_i$, respectively).  To examine the performance of measuring the $ E_{ph} $
	spectrum, the primary X-ray  energy in a normal distribution with $
	\mu=621\ eV $ (average) and $ \sigma=0.1\ eV $ (deviation) was
	simulated. The estimated parameters ($ \hat{\mu}\ \text{and}\
	\hat{\sigma} $) can be obtained by fitting the decomposed signal
	with a normal distribution. Since $ \hat{\mu} $ can be corrected
	by using another advanced spectrometer,  $ \hat{\sigma}-\sigma $ might be
	the figure of merit (FOM) for evaluating spectrometer
	performance.

\begin{figure}[htp]
		\centering
		\includegraphics[width=70mm]{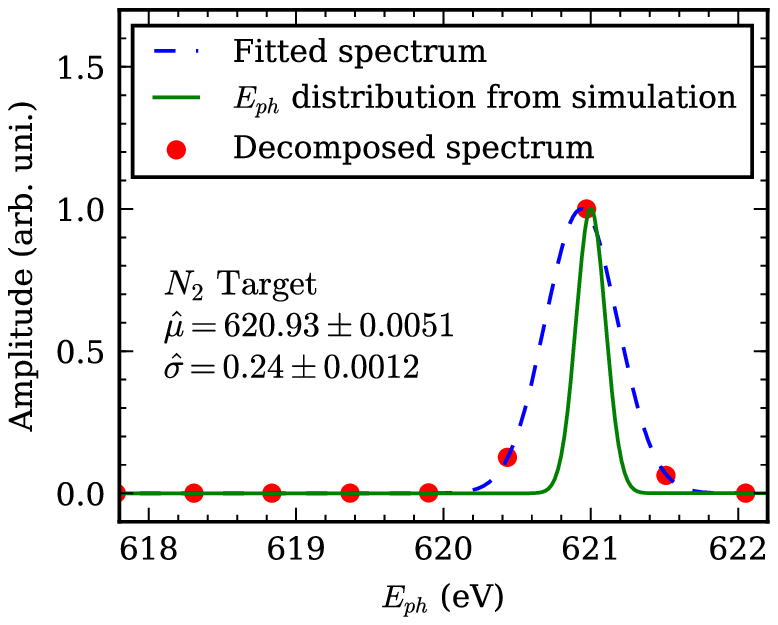}
		\qquad
		\includegraphics[width=70mm]{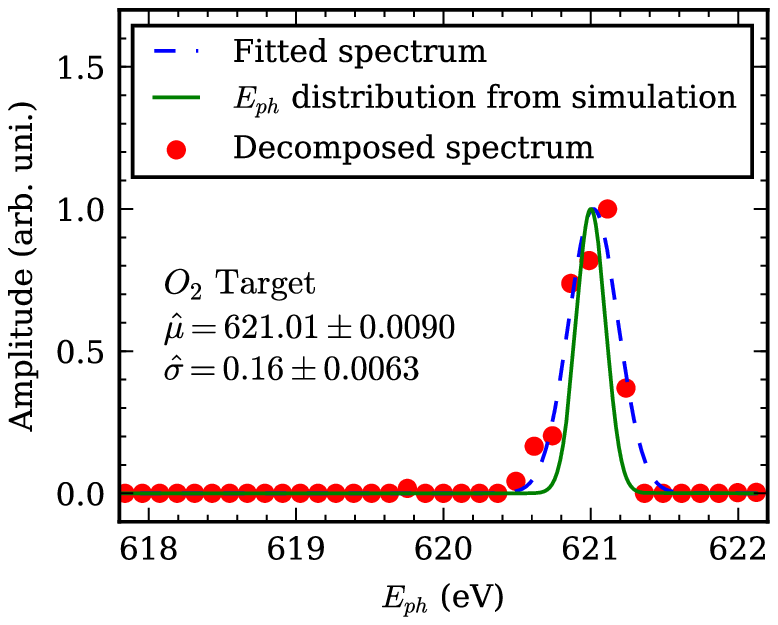}
		\caption{Simulation result for the $ N_2$ target (left) and the  $ O_2$
			target (right)  with a sampling time interval of $\Delta t_s=100\
			(ps)$ and a resampling time interval of  $\Delta t_i=50\ (ps)$.}
		\label{fig:spec-gas}
	\end{figure}

	According to Eq. \ref{eq:binding energy}, a target gas with a higher
	binding energy means a lower photoelectron kinetic energy, which results in longer
	signal duration for a given drift length and furthermore a
	higher resolution when other parameters remain the same. Both nitrogen and
	oxygen are acceptable and their binding energies are $\sim$403  and  $\sim$ $540.43\ eV $, respectively \cite{TRZHASKOVSKAYA2001}. 
	The result of the $E_{ph}$ spectrum measurement simulations for these
	two kinds of target gases are demonstrated in Fig.
	\ref{fig:spec-gas}. According to the simulation, $ \hat{\sigma} $
	for the $ O_2 $ target is much smaller than that for $ N_2 $. Therefore, $ O_2
	$ is more suitable for the  $ E_{ph} $ spectrum measurement.

	\begin{figure}[htp]
		\centering
		\includegraphics[width=70mm]{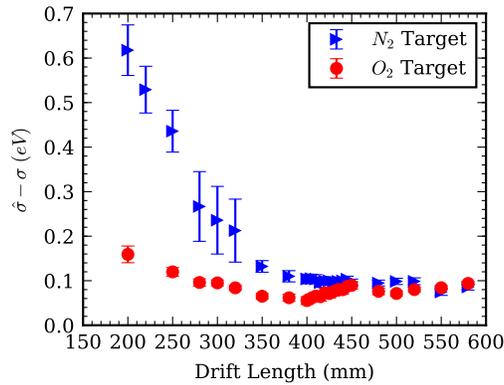}
		\caption{Deviation of $ \hat{\sigma} $ with changing drift
			distance.}
		\label{fig:spec-driftdistance}
	\end{figure}

	\begin{figure}[H]
		\begin{minipage}[t]{0.4\textwidth}
			\centering
			\includegraphics[width=70mm]{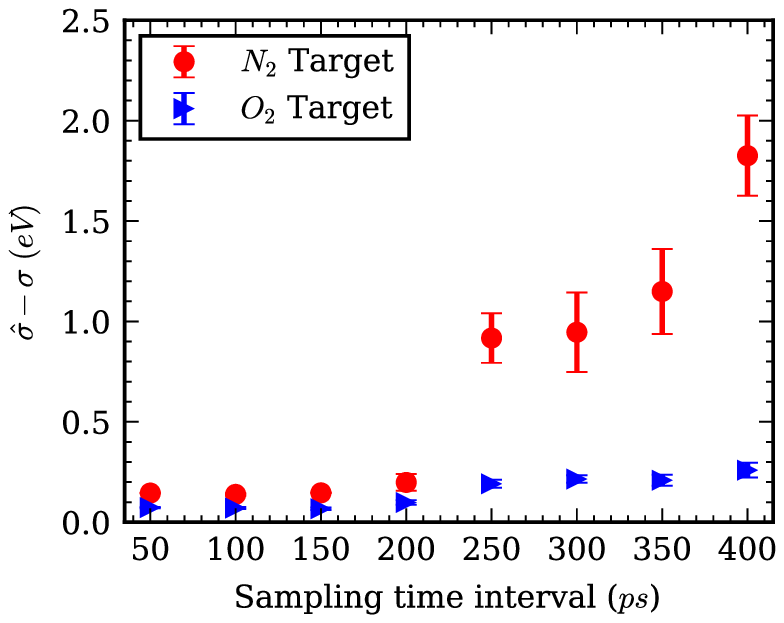}
			\caption{Deviation of $ \hat{\sigma} $ with changing 
				sampling time interval $ \Delta t_s $.}
			\label{fig:spec sam-binwidth}
		\end{minipage}
		\hspace{60pt}
		\begin{minipage}[t]{0.4\textwidth}
			\centering
			\includegraphics[width=70mm]{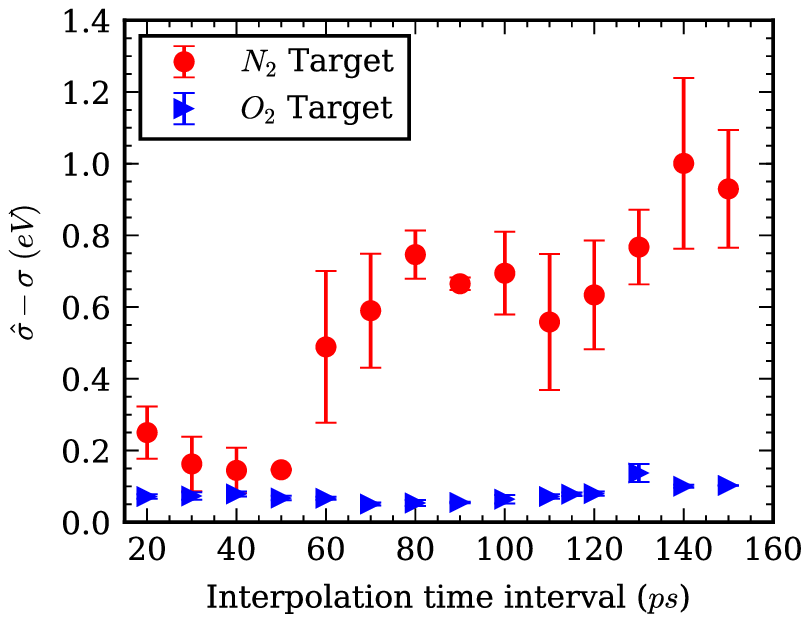}
			\caption{Deviation of $ \hat{\sigma} $ with changing
				interpolation time interval $ \Delta t_i $.}
			\label{fig:spec inter-binwidth}
		\end{minipage}
	\end{figure}
	
		A longer drift length converts small energy differences of electrons
	into  measurable time and helps to extend the signal duration for
	digitizer sampling.  The simulation result for the dependence
	of drift distance $ L $ is  reported in Fig.
	\ref{fig:spec-driftdistance}.  From the result, it can be
	inferred that $ L $ should be $\ge$ $350\ mm$ for $ N_2 $. In contrast, for the
	$ O_2 $ target, the smallest deviation of $ \hat{\sigma} $ is
	located at $L\sim 400\ mm$, which can be understood
	because a longer drift length means fewer detected photoelectrons
	while a shorter drift length results in a shorter flight time. In short, the
	$ O_2 $ target is more suitable than $ N_2 $ and $ L=400\ mm $
	for the $ O_2 $ target is the optimal design.
	
	Nowadays, analog-to-digital converter sampling rates are as high as  $30\
	Gs$\cite{Zhu2017A}. Generally, $t_s$ from $50$ to $400\ ps$ is
	practicable and acceptable. By changing $\Delta t_s$ and $\Delta
	t_i$, the variation of $ \hat{\sigma} $ can be demonstrated, as shown in Figs.
	\ref{fig:spec sam-binwidth} and \ref{fig:spec
		inter-binwidth}.
	According to Figs. \ref{fig:spec sam-binwidth} and
	\ref{fig:spec inter-binwidth}, $ \hat{\sigma} $ for the $ O_2 $
	target is better than that for $ N_2 $ when both $ \Delta t_s $
	and $ \Delta t_i $ are the same. Moreover, for the $ O_2 $ target, the
	design requirement can be satisfied as long as $\Delta t_s $
	and $\Delta t_i $ are within the simulation range (i.e., $ \Delta t_i <
	\Delta t_s$ is required).
	
	In addition, the same method for energy spectrum reconstruction
	was used to separate photons with two energies from the same X-ray
	FEL pulse under optimized conditions ($ E_{ph1}=621\ eV$, $
	E_{ph2}=621\ { eV} +\Delta E $, and $ \Delta t_s=150\ ps $ for an $ O_2 $ 
	target). The scan was performed by changing $ \Delta E $ and the
	two energy peaks were obvious when $ \Delta E \geq0.3\ eV $,
	giving an energy resolution of $ 0.3\ eV $
	
	\section{Discussion  and conclusion }
	Based on the previous design concept of an e-TOF instrument \cite{Allaria2014Control} and the numerical optimization by using simulation tools, we have validated the
	feasibility of an e-TOF-based polarimeter and spectrometer for an 
	X-ray FEL user facility. The polarization resolution is $ 0.5\%$
	and the  photon's energy resolution is $  0.3\ eV $ for the
	optimized design and the following requirements are obtained: 
	\begin{enumerate}
		\item The target region's pressure must be in the range from $ 0.1 $
		to $ 1.0\ Pa $ and the drift region's pressure  must be $<$ $ 0.1\ Pa $.
		\item  The drift distance $ L $  must be $>$400 mm.
		\item  The diameter of the detectors must be $>$ $30\ mm$.
		\item  The remanent magnetic field must be controlled to be $<$ $ 0.2\ gauss $.
		\item The sampling time interval must be in the range from $50 $ to
		$400\ ps$ and the interpolation time interval must be in the range from
		$20 $ to $140\ ps$.
	\end{enumerate}
	
	There are still a few conditions to be considered for further
	study, such as the influence on $ P_{lin} $ from photoelectrons
	from $\emph{p}$ subshells and the smaller effect of Auger
	electrons on the polarity measurement.
	
	\acknowledgments
	This work is supported by the Fundamental Research Funds for
	the Central Universities (Grant No. xjj2017109), the National Natural Science Foundation of China (Grant No. 11775293), the Natural Science
	Fundamental Research Plan of Shaanxi Province (Grant No. 2016JM1019), the China Association for Science and Technology, and the Ten Thousand Talents Program.
	\raggedright

\end{document}